\documentclass[runningheads]{llncs}
\usepackage{tikz}
\usetikzlibrary{patterns}

\usepackage{graphicx}
\usepackage{comment}
\usepackage{tocbibind}
\usepackage{longtable} 
\usepackage{array}
\usepackage{booktabs}
\usepackage{pgfplots}
\pgfplotsset{compat=1.18}
\usepackage{amsmath}
\usepackage{amssymb}
\usepackage{algorithm} 
\usepackage{algpseudocode}
\usepackage{multirow}
\usepackage{hyperref}
\hypersetup{
  colorlinks=false,
  pdfborder={0 0 0}
}
\usepackage[numbers, sort&compress]{natbib}

\newcommand{\repeatthanks}{\textsuperscript{\thefootnote}}

\begin{document}
\title{Adaptive AI Task Partitioning and Safe Offloading in Heterogeneous Edge-Cloud Continuum} 
\titlerunning{Adaptive AI Task Partitioning and Safe Offloading}

\author{Akuen~Akoi~Deng\thanks{These authors contributed equally to this work.} \orcidID{0009-0007-6228-3340} \and
Eimantas~Butkus\repeatthanks \orcidID{0009-0001-5647-0779} \and
Alfreds~Lapkovskis\orcidID{0009-0003-4424-949X} \and
Praveen~Kumar~Donta\orcidID{0000-0002-8233-6071}}
\authorrunning{A. Deng et al.}
%
\institute{Department of Computer and Systems Sciences, Stockholm University, Borgarfjordsgatan 12, Stockholm, Sweden \\
\email{\{akuiendng, butkus505\}@gmail.com \\ \{alfreds.lapkovskis, praveen\}@dsv.su.se}}

\maketitle              
\begin{abstract}
In recent years, the use of artificial intelligence on resource-constrained IoT devices has grown significantly. However, existing approaches to AI task partitioning and offloading across the edge-cloud continuum typically rely on static methods that ignore runtime dynamics. Furthermore, they are often evaluated in simulated environments rather than on real hardware. To address this gap, we propose a framework that dynamically splits neural network layers across the heterogeneous continuum. The framework profiles the model at startup, measures network link conditions between nodes, and periodically re-evaluates the partition to adapt to environmental changes. We created a physical testbed comprising a Raspberry Pi edge device, a laptop fog, and a high-performance desktop PC as the cloud. We evaluated the framework over three widely adopted convolutional neural networks: VGG16, AlexNet, and MobileNetV2. Our results show that the framework achieves reductions in energy and end-to-end latency of 27.09--35.82\% and 6.34--22.92\%, respectively, compared to a static partitioning baseline. These findings confirm the superiority of adaptive to static partitioning.

\end{abstract}
\section{Introduction}
In recent years, the widespread adoption of Internet-of-Things (IoT) devices
has led to unprecedented growth in the volume, velocity, and variety of data
generated~\cite{edquist2021internet,chen2022task,8089336}. These devices range
from smartphones and wearables to RFID readers and tablets, increasingly
supporting applications with strict requirements for latency, energy efficiency,
and computational workload~\cite{8528362,chen2022task}, such as smart homes,
robotics, virtual reality, autonomous driving, and M2M
communications~\cite{8089336,feng2022computation}. Simultaneously, rapid
advancements in artificial intelligence (AI), specifically deep neural networks
(DNNs), have enabled fast and sophisticated data-driven decision-making across
diverse fields~\cite{8697857,dargan2020survey}.

Traditionally, AI models have been deployed and executed in a centralized
environment, such as the cloud, which provides abundant computational resources.
However, this approach introduces significant limitations when applied to IoT
systems~\cite{chen2022task}. In an IoT environment, network latency, privacy
concerns, bandwidth constraints, and poor connectivity can significantly degrade
real-time system performance~\cite{chen2024edgeci}. This has driven
the research toward a computing continuum spanning cloud, edge,
fog, and IoT devices~\cite{computers12100198}, where computing is increasingly
distributed across these heterogeneous tiers to meet the
demands of modern AI applications~\cite{ye2026nesy}. This approach allows
some parts of the AI workload to run on edge devices alongside the central cloud
servers, a model known as edge-cloud AI, promising to combine the low-latency
benefits of edge computing with the scalability of the
cloud~\cite{8123913,zhang2020implementation,8089336}.

However, designing adaptive and efficient AI systems for resource-constrained edge devices
remains challenging due to the inherent heterogeneity of the edge-cloud
environment, where devices vary widely in computational capability, energy
budget, and network conditions~\cite{doan2024optimizing,11270629}. To address this,
researchers have explored distributed training, deployment, and inference of
AI models across heterogeneous
nodes~\cite{doan2024optimizing,li2020pytorch,langer2020distributed}.

In distributed inference, AI workloads are split across edge and cloud devices,
allowing latency-critical tasks to execute close to the data source while more
computationally intensive operations are offloaded to the
cloud~\cite{zhao2018deepthings,zhao2025adapcp}. However, most current approaches
rely on static deployment with predefined inference pipelines that do not adapt
to runtime changes. Factors such as fluctuating bandwidth, workload contention,
and node mobility make fixed partitioning decisions inefficient in dynamic
environments~\cite{chen2022task}, frequently resulting in performance
degradation. This motivates the need for an adaptive partitioning and offloading
framework that dynamically distributes AI workloads across edge-cloud resources
while satisfying stringent latency and energy constraints.

The literature on adaptive AI task partitioning and offloading in heterogeneous
edge-cloud networks has expanded significantly in recent
years~\cite{li2025adaptive,shen2025reinforcement,chen2022task}. Proposed
approaches include reinforcement learning-based layer partitioning
strategies~\cite{shen2025reinforcement,10.1145/3773274.3774271}, collaborative
caching mechanisms~\cite{fang2022ai}, and edge-end joint parallel inference
architectures~\cite{zhao2025adapcp}. However, a key limitation of many existing
solutions is their reliance on simulated environments that assume ideal network
conditions~\cite{li2025adaptive,10.1145/3773274.3774271,electronics14081647},
leaving the practical performance of these methods in real-world deployments
insufficiently explored.

Li et al.~\cite{li2025adaptive} proposed a framework that jointly optimises
fine-grained DNN layer partitioning and magnitude-based channel pruning to
reduce inference latency, using an LSTM-based controller trained with a
policy gradient algorithm and a delay-aware reward function. Despite promising
results, the evaluation was conducted entirely through simulation in an MEC
environment.

Studies conducted in real environments, e.g., \cite{fang2022ai}, tend to focus
on metrics such as task completion time, while overlooking energy consumption and
computational load balancing. Fang et al.~\cite{10.1145/3551638} proposed
EdgeDI, which compresses models via structured pruning and a lightweight
attention-based convolutional block, then partitions feature-map workloads
across industrial IoT devices using an execution-time-aware scheme that accounts for
both computation capability and network bandwidth. While EdgeDI achieves
notable inference speedups, the work focuses exclusively on inference
performance, leaving energy consumption unaddressed.

These examples reflect a consistent trend in the literature: partitioning and
offloading frameworks are predominantly evaluated in simulation, and energy
consumption is rarely a primary design objective in heterogeneous edge-cloud
environments.

This work addresses both gaps through real-world experimentation on
resource-constrained hardware, proposing an adaptive partitioning and offloading
framework that minimizes energy consumption without violating real-time latency
constraints. Our contributions are threefold:
\begin{itemize}
    \item We propose an adaptive partitioning and offloading framework for
    DNN inference across a heterogeneous edge-cloud continuum.

    \item We build a fully functional three-tier testbed comprising a Raspberry
    Pi edge device, a laptop fog node, and a high-performance PC cloud node,
    enabling evaluation of vertical offloading strategies on real hardware.

    \item We conduct real-world experiments over three widely adopted convolutional neural network (CNN) models
    (VGG16, AlexNet, and MobileNetV2), collecting and evaluating hardware-measured
    energy and latency metrics across single-device, static partitioning, and
    adaptive partitioning configurations.
\end{itemize}

The remainder of this paper is organized as follows. Section~\ref{sec:framework}
presents the proposed adaptive partitioning framework. Section~\ref{sec:results}
describes the experimental setup and discusses the results. Section~\ref{sec:conclusions}
concludes the paper.

\section{Proposed Framework} \label{sec:framework}

We assume a continuum environment with an edge, fog, and a cloud node. In this architecture, inference is initiated on the edge device, intermediate DNN activations are deployed to the fog, and the remaining computations are performed on the cloud. This allows partitioning schemas with $0,\dots,i$ layers on the edge, $i+1,\dots,j$ layers on the fog, and $j+1,\dots N-1$ layers on the cloud, where $1\leq i < j< N$ are layer indices.

The proposed framework dynamically partitions and offloads CNN inference workloads across nodes to minimize energy consumption without exceeding latency limits.
At runtime, the framework periodically evaluates the current partition against available resources. It only reconfigures the workload if real-time computation and communication measurements predict a more optimal split. 

The framework follows a five-stage algorithmic pipeline: (i) offline profiling, (ii) two-point link probing, (iii) candidate split estimation, (iv) best candidate search, and (v) adaptive distributed inference scheduling. The following sections detail these stages.

\subsection{Offline Profiling}
The offline profiler runs once before any partitioning decision is made, producing
two lookup tables used throughout the framework. The first is an activation size
table, which records the size of the intermediate tensor in bytes at every feature
boundary. For a candidate split $(i, j)$, this gives the number of bytes the
Raspberry Pi must transmit to the laptop after layer $i$, and the laptop must
transmit to the PC after layer $j$. The second is a relative compute weight table,
where each layer is assigned a normalized fraction of the total inference work
based on a single measured execution. These weights serve as a stable reference
that allows runtime measurements from a small number of probe splits to be scaled
and generalized to estimate the latency and energy of any unseen partition. The offline profiling algorithm is shown in Algorithm \ref{alg:profile}.

\begin{algorithm}[H]
\caption{Offline profiling of a neural network $\mathcal{N}$}
\label{alg:profile}
\begin{algorithmic}[1]
\Require Pretrained network $\mathcal{N}$ with ordered feature children
         $\mathcal{F} = (f_0, f_1, \ldots, f_{N-1})$ and classifier head $H$
\Ensure  Per-layer activation sizes $B$ and relative compute weights $W$
\State $x \gets \text{randn}(1, 3, 224, 224)$
\For{$k \gets 1$ to $3$} \Comment{warmup}
    \State $x' \gets H(\mathcal{F}(x))$
\EndFor
\State $x \gets \text{randn}(1, 3, 224, 224)$
\For{$i \gets 0$ to $N-1$}
    \State $t_0 \gets \text{now}()$
    \State $x \gets f_i(x)$
    \State $T[i] \gets \text{now}() - t_0$
    \State $B[i] \gets \text{numel}(x) \cdot \text{elementSize}(x)$
\EndFor
\State $t_0 \gets \text{now}()$;\; $\_ \gets H(x)$;\; $T[N] \gets \text{now}() - t_0$
\State $W[k] \gets T[k] \,/\, \sum_{j} T[j]$ \quad for all $k \in \{0, \ldots, N\}$
\State \Return $B, W$
\end{algorithmic}
\end{algorithm}

\subsection{Two-Point Link Probing}

With activation sizes known, the framework must also estimate the communication
cost between nodes. The two-point link probe achieves this by transmitting two
payloads of contrasting sizes $s_1$ and $s_2$ across each hop, repeating each
transmission multiple times and averaging the round-trip times to reduce
short-term timing noise. The round-trip time for a payload of size $s$ on hop
$h$ is modeled as

\begin{equation}\label{eq:omega}
    rtt_h(s) = \omega + \frac{s}{\beta},
\end{equation}
where $\omega$ is the fixed communication overhead and $\beta$ is the effective
throughput in bytes per second. Given the averaged round-trip times $\tau[s_1]$
and $\tau[s_2]$, the throughput is estimated as
\begin{equation}\label{eq:beta1}
    \beta = \frac{s_2 - s_1}{\tau[s_2] - \tau[s_1]}
\end{equation}
and the fixed overhead as
\begin{equation}\label{eq:omega1}
    \omega = \max\left(0, \tau[s_1] - \frac{s_1}{\beta}\right).
\end{equation}
The resulting pair $(\omega, \beta)$ provides a concrete link model that the
framework uses to predict the transfer time of any activation tensor at a given
split point, as formalised in Algorithm~\ref{alg:linkprobe}.

\begin{algorithm}[H]
\caption{Two-point link probe}
\label{alg:linkprobe}
\begin{algorithmic}[1]
\Require Hop $h$ with round-trip function $\text{rtt}_h(s)$;\; sizes
         $s_1 \ll s_2$;\; repeat count $r$
\Ensure  Link model $(\omega, \beta)$ where $\omega$ is fixed overhead and
         $\beta$ is throughput in bytes/second
\For{$s \in \{s_1, s_2\}$}
    \State $\tau[s] \gets \text{mean}\big(\{\text{rtt}_h(s) : k = 1,\ldots,r\}\big)$
\EndFor
\If{$\tau[s_2] \le \tau[s_1]$} \Return previous model \Comment{malformed probe; keep stale values}
\EndIf
\State compute $\beta$ using Eq.(\ref{eq:beta1})
\State $\omega \gets \max\!\big(0,\; \tau[s_1] - s_1/\beta\big)$
\State \Return $(\omega, \beta)$
\end{algorithmic}
\end{algorithm}

\subsection{Candidate Split Latency and Energy Estimation}

For a candidate split $(i, j)$, where $i$ is the index of the last feature layer
executed on the edge node and $j$ is the last feature layer on the fog node, the
estimator predicts end-to-end latency and energy consumption before the split is
actually run. Latency is computed as the sum of three per-node computation times
and two inter-node transfer times, derived from the profiled compute weights, the
per-node execution rates, and the fitted link models. Energy is estimated per
node by multiplying each node's compute time by its corresponding power or energy
rate, where the edge node applies a fixed power model, and the fog and cloud nodes
use empirical rates fitted from previous runs. The total system energy is the sum
of all three contributions. Since these are estimates rather than measurements,
any discrepancy between the predicted and observed values is used to refine the
per-node rates in the next re-evaluation cycle, as formalized in
Algorithm~\ref{alg:estimate}.

\begin{algorithm}[H]
\caption{Estimate latency and energy of a candidate split $(i, j)$}
\label{alg:estimate}
\begin{algorithmic}[1]
\Require Candidate $(i, j)$;\; per-layer compute weights $W$;\; per-node
         speeds $(\sigma_{\text{Pi}}, \sigma_{\text{Lap}}, \sigma_{\text{PC}})$;\;
         per-node energy rates $(\rho_{\text{Lap}}, \rho_{\text{PC}})$;\;
         link models $(\omega_{pl}, \beta_{pl})$ and $(\omega_{lp}, \beta_{lp})$;\;
         payload table $B$
\State $w_{\text{Pi}} \gets \sum_{k=0}^{i} W[k]$
\State $w_{\text{Lap}} \gets \sum_{k=i+1}^{j} W[k]$
\State $w_{\text{PC}} \gets \sum_{k=j+1}^{N} W[k]$
\State $t_{\text{Pi}} \gets \sigma_{\text{Pi}} \cdot w_{\text{Pi}}$;\;
       $t_{\text{Lap}} \gets \sigma_{\text{Lap}} \cdot w_{\text{Lap}}$;\;
       $t_{\text{PC}} \gets \sigma_{\text{PC}} \cdot w_{\text{PC}}$
\State $t_{pl} \gets \omega_{pl} + B[i] / \beta_{pl}$
\State $t_{lp} \gets \omega_{lp} + B[j] / \beta_{lp}$
\State $\hat{L} \gets t_{\text{Pi}} + t_{\text{Lap}} + t_{\text{PC}} + t_{pl} + t_{lp}$
\State $\hat{E}_{\text{Pi}} \gets P_{\text{Pi}} \cdot t_{\text{Pi}}$
       \Comment{$P_{\text{Pi}}$: fixed Pi power model (12\,W)}
\State $\hat{E}_{\text{Lap}} \gets \rho_{\text{Lap}} \cdot t_{\text{Lap}}$;\;
       $\hat{E}_{\text{PC}} \gets \rho_{\text{PC}} \cdot t_{\text{PC}}$
\State $\hat{E}_{\text{tot}} \gets \hat{E}_{\text{Pi}} + \hat{E}_{\text{Lap}} + \hat{E}_{\text{PC}}$
\State \Return $(\hat{L},\; \hat{E}_{\text{Pi}},\; \hat{E}_{\text{tot}})$
\end{algorithmic}
\end{algorithm}

\subsection{Best Candidate Split Search}
The candidate search evaluates every valid split pair $(i, j)$, where validity
requires each node to execute at least one layer, and returns the candidate that
minimizes the objective score. For each candidate, the estimator is called to
obtain the predicted latency and energies, after which two filters are applied.
The first rejects any candidate whose predicted latency exceeds the deadline
$L_{\max}$, and the second rejects candidates whose normalized score exceeds
that of the static baseline, guaranteeing the adaptive framework never produces
a result worse than the baseline it is intended to improve upon. Candidates that
pass both filters are scored using the weighted sum

\begin{equation}
S(i,j) \triangleq w_{\text{end}} \frac{\hat{E}_{\text{end}}(i,j)}{n_{\text{end}}} +
w_{\text{tot}} \frac{\hat{E}_{\text{tot}}(i,j)}{n_{\text{tot}}} +
w_{\text{lat}} \frac{\hat{L}(i,j)}{n_{\text{lat}}},
\label{eq:weighted_objective}
\end{equation}
where $w_{\text{end}}$, $w_{\text{tot}}$, and $w_{\text{lat}}$ are user-defined
weights, and the normalization anchors $n$ are average values measured from
probe splits run at startup. Here, $\hat{E}_{\text{end}}(i,j)$ and
$\hat{E}_{\text{tot}}(i,j)$ denote the predicted edge node and total system
energy respectively, and $\hat{L}(i,j)$ denotes the predicted end-to-end
latency for split $(i,j)$, as estimated by Algorithm~\ref{alg:estimate}. Normalization makes the score dimensionless and
ensures each weight exerts a comparable influence regardless of the absolute
magnitude of the underlying quantities, as formalized in
Algorithm~\ref{alg:findbest}.

\begin{algorithm}[t]
\caption{Find best candidate split}
\label{alg:findbest}
\begin{algorithmic}[1]
\Require Rates and link models (as in Alg.~\ref{alg:estimate});\;
         weights $\mathbf{w} = (w_{\text{Pi}}, w_{\text{tot}}, w_{\text{lat}})$;\;
         normalization anchors $\mathbf{n} = (n_{\text{Pi}}, n_{\text{tot}}, n_{\text{lat}})$;\;
         baseline score $S^\star$;\; deadline $L_{\max}$;\;
         minimum Pi layers $m$;\; current split $c$
\State $(s^\star_{\text{best}},\, S_{\text{best}}) \gets (\textsc{None},\, +\infty)$
\For{$(i, j) \in \{(i, j) : m\!-\!1 \le i < j < N\}$}
    \If{$(i, j) = c$} \textbf{continue} \Comment{exclude currently-running split}
    \EndIf
    \State $(\hat{L},\hat{E}_{\text{Pi}},\hat{E}_{\text{tot}}) \gets
           \textsc{Estimate}(i, j, \ldots)$ \Comment{Alg.~\ref{alg:estimate}}
    \If{$L_{\max} > 0$ \textbf{and} $\hat{L} > L_{\max}$} \textbf{continue}
    \Comment{deadline pre-filter}
    \EndIf
    \State $S \gets w_{\text{Pi}} \cdot \dfrac{\hat{E}_{\text{Pi}}}{n_{\text{Pi}}}
                + w_{\text{tot}} \cdot \dfrac{\hat{E}_{\text{tot}}}{n_{\text{tot}}}
                + w_{\text{lat}} \cdot \dfrac{\hat{L}}{n_{\text{lat}}}$
    \If{$S > S^\star$} \textbf{continue} \Comment{must beat static baseline}
    \EndIf
    \If{$S < S_{\text{best}}$}
        \State $(s^\star_{\text{best}},\, S_{\text{best}}) \gets ((i, j),\, S)$
    \EndIf
\EndFor
\State \Return $s^\star_{\text{best}}$
\end{algorithmic}
\end{algorithm}

\subsection{Adaptive Distributed Inference Scheduling}

The adaptive scheduler orchestrates the full runtime loop across two phases,
with initialization handled in phase 1 and steady-state operation in phase 2.

In phase 1a, the scheduler runs a user-defined initial split for a fixed number
of inferences to record the nodes' latency and energy consumption, establishing
the baseline threshold that subsequent candidates must improve upon. In phase 1b,
three additional probe splits are run automatically, selected to expose
edge-heavy, balanced, and cloud-heavy workload distributions, giving the estimator
sufficient data to infer per-node execution rates across a wide operating range.
In phase 1c, the collected baseline and probe measurements are used to fit the
per-node compute rates, probe both network links, and select the best starting
split by evaluating all candidates against the objective score defined in
Eq.~(\ref{eq:weighted_objective}). Energy terms are weighted more heavily than
latency, with the edge node energy weight ranging from 0.6 to 0.9, total system
energy from 0.2 to 0.3, and latency from 0.1 to 0.3, reflecting the framework's
primary objective of reducing energy consumption without violating latency
constraints.

In phase 2, the scheduler enters steady-state operation and runs the current
split for $R_{\text{steady}}$ inferences per window. At the end of each window,
it refits the per-node compute rates using both the phase 1 data and the most
recent window combined, re-probes both network links, and re-runs the candidate
search with the updated parameters. A switch is made if the new candidate
improves the objective score by at least 3\%. If the latency deadline $L_{\max}$
is violated during a window, the switch is forced regardless of the improvement
margin; if no better candidate exists under a deadline violation, the scheduler
falls back to the static baseline $c_0$ as the safest known configuration. This
periodic re-evaluation is what makes the framework adaptive, as changes in
network bandwidth or node compute load are captured by the re-probing and
re-fitting steps and reflected in the next candidate selection, as formalized in
Algorithms~\ref{alg:scheduler} and~\ref{alg:scheduler2}.

\begin{algorithm}[t]
\caption{Adaptive distributed inference scheduler -- Initialization}
\label{alg:scheduler}
\begin{algorithmic}[1]
\Require Model $\mathcal{N}$;\; initial split $c_0$;\;
         weights $\mathbf{w}$;\; deadline $L_{\max}$;\;
         $R_{\text{profile}}$ baseline runs;\;
         $R_{\text{probe}}$ runs per probe split;\;
         $R_{\text{steady}}$ runs per re-evaluation window;\;
         warmup $k_{\text{warm}}$;\;
         switch improvement threshold $\theta$
\Statex
{\textbf{Phase 1a:} baseline run (defines threshold to beat)}
\State $(B, W) \gets \textsc{Profile}(\mathcal{N})$ \Comment{Alg.~\ref{alg:profile}}
\State $\mathcal{D}_{\text{base}} \gets \emptyset$;\; $c \gets c_0$
\For{$r \gets 1$ to $R_{\text{profile}}$}
    \State $s \gets \textsc{RunInference}(c)$
    \If{$r > k_{\text{warm}}$} $\mathcal{D}_{\text{base}} \gets \mathcal{D}_{\text{base}} \cup \{s\}$
    \EndIf
\EndFor
\State $(b_{\text{Pi}}, b_{\text{tot}}, b_{\text{lat}}) \gets
        \text{mean energies and latency over } \mathcal{D}_{\text{base}}$
\Statex
{\textbf{Phase 1b:} probe reference splits to ground per-layer rates}
\State $\mathcal{P} \gets \textsc{ProbeSplits}(N, m)$
       \Comment{3 splits at fifths of the feature range}
\State $\mathcal{D}_{\text{probe}} \gets \emptyset$
\ForAll{$p \in \mathcal{P} \setminus \{c_0\}$}
    \For{$r \gets 1$ to $R_{\text{probe}}$}
        \State $s \gets \textsc{RunInference}(p)$
        \If{$r > k_{\text{warm}}$}
            $\mathcal{D}_{\text{probe}} \gets \mathcal{D}_{\text{probe}} \cup \{s\}$
        \EndIf
    \EndFor
\EndFor
\Statex
{\textbf{Phase 1c:} fit rates, probe links, choose starting split}
\State $\mathbf{n} \gets \text{mean energies and latency over } \mathcal{D}_{\text{probe}}$
       \Comment{normalization independent of $c_0$}
\State $S^\star \gets w_{\text{Pi}} (b_{\text{Pi}}/n_{\text{Pi}})
                    + w_{\text{tot}} (b_{\text{tot}}/n_{\text{tot}})
                    + w_{\text{lat}} (b_{\text{lat}}/n_{\text{lat}})$
\State $\sigma, \rho \gets \textsc{FitRates}(\mathcal{D}_{\text{base}} \cup \mathcal{D}_{\text{probe}},\, W)$
\State $(\omega_{pl}, \beta_{pl}), (\omega_{lp}, \beta_{lp}) \gets \textsc{LinkProbe}()$
       \Comment{Alg.~\ref{alg:linkprobe}}
\State $c \gets \textsc{FindBest}(\sigma, \rho, \omega, \beta, \mathbf{w}, \mathbf{n}, S^\star, L_{\max}, m, \textsc{None})$
       \Comment{Alg.~\ref{alg:findbest}}
\end{algorithmic}
\end{algorithm}
 
\begin{algorithm}[t]
\caption{Adaptive distributed inference scheduler -- Steady state}
\label{alg:scheduler2}
\begin{algorithmic}[1]
\Require Current split $c$;\; baseline split $c_0$;\;
         weights $\mathbf{w}$;\; deadline $L_{\max}$;\;
         $R_{\text{steady}}$ runs per re-evaluation window;\;
         warmup $k_{\text{warm}}$;\;
         switch improvement threshold $\theta$;\;
         Phase-1 data $\mathcal{D}_{\text{phase1}} = \mathcal{D}_{\text{base}} \cup \mathcal{D}_{\text{probe}}$
\Statex
{\textbf{Phase 2:} steady state with periodic re-evaluation}
\While{budget not exhausted}
    \State $\mathcal{W} \gets \emptyset$
    \For{$r \gets 1$ to $R_{\text{steady}}$}
        \State $s \gets \textsc{RunInference}(c)$
        \If{$r > k_{\text{warm}}$}\;
            $\mathcal{W} \gets \mathcal{W} \cup \{s\}$
        \EndIf
    \EndFor
    \State $\bar{L} \gets \text{mean latency over } \mathcal{W}$
    \State $\sigma, \rho \gets \textsc{FitRates}(\mathcal{D}_{\text{phase1}} \cup \mathcal{W},\, W)$
           \Comment{Phase-1 data keeps rates grounded}
    \State $(\omega, \beta) \gets \textsc{LinkProbe}()$
    \State $c' \gets \textsc{FindBest}(\sigma, \rho, \omega, \beta, \mathbf{w}, \mathbf{n}, S^\star, L_{\max}, m, c)$
    \State $S_c, S_{c'} \gets \textsc{Score}(c, \ldots), \textsc{Score}(c', \ldots)$
    \State $\Delta \gets (S_c - S_{c'}) / S_c$ \Comment{relative improvement}
    \Statex
    \State $\text{deadlineHit} \gets (L_{\max} > 0 \land \bar{L} > L_{\max})$
    \If{$\text{deadlineHit} \land c' \ne c$}
        \State $c \gets c'$ \Comment{\textsc{forced} switch (deadline violation)}
    \ElsIf{$c' \ne c \land \Delta \ge \theta$}
        \State $c \gets c'$ \Comment{\textsc{normal} switch}
    \ElsIf{$\text{deadlineHit} \land c \ne c_0$}
        \State $c \gets c_0$ \Comment{fall back to static baseline}
    \EndIf
\EndWhile
\end{algorithmic}
\end{algorithm}

\section{Results and Discussion} \label{sec:results}
This section presents the quantitative evaluation of the proposed adaptive
partitioning framework against single-device and static partitioning baselines
across three CNN models.

\subsection{Experimental Setup}
Experiments were conducted on a three-tier heterogeneous testbed representing
a typical IoT edge-cloud hierarchy. The edge device is a Raspberry Pi 4
(4-core Arm Cortex-A76, 2.4 GHz, 8 GB LPDDR4X), representing a
resource-constrained IoT node. The fog node is a laptop (Intel Core
i7-10510U, 16 GB DDR4), and the cloud node is a desktop PC equipped with an
NVIDIA RTX 4070 Ti GPU (7680 CUDA cores, 40.09 FP32 TFLOPS, 32 GB DDR5,
12 GB VRAM GDDR6). Nodes were interconnected via Tailscale VPN with ZeroMQ
handling inter-node messaging. Network constraints and bottlenecks were
simulated using Tailscale's traffic throttling capabilities.

Three widely benchmarked CNN models were evaluated: VGG16 (138M parameters,
528 MB), AlexNet (61M parameters, 234 MB), and MobileNetV2 (2.2M parameters,
8.8 MB), spanning a broad range of complexity. To isolate computational
workload from dataset variability, randomly generated dummy tensors of shape
$1 \times 3 \times 224 \times 224$ matching each model's input dimensions
were used as inference inputs. Models were implemented in PyTorch.

Energy consumption was measured independently per node: the edge node applied
a constant 12\,W power model multiplied by measured compute time, the fog
node read package energy deltas via Intel RAPL through the Linux powercap
interface, and the cloud node integrated instantaneous GPU power readings from
NVIDIA NVML over the compute window. Latency was recorded using wall-clock timing. Two primary metrics are reported: total end-to-end inference latency
(ms) and total energy consumption (J) across all active nodes.

Each configuration was executed over 500 inference passes, repeated ten times
and averaged to ensure measurement stability. We compare three execution
strategies: (i) single-device baseline on each node independently, (ii) static
partitioning with a fixed split, approximating equal workload thirds across the
three nodes, and (iii) the proposed adaptive partitioning algorithm, which
dynamically determines split points based on runtime conditions.

The implementation is publicly available on GitHub.\footnote{\url{https://github.com/Akuien/DNN-partitioning-and-Oflloading-framework-REAP}}

\subsection{Single-Device Baselines}

Table~\ref{tab:single_device_baseline} reports the average per-inference latency
and energy consumption for each model running entirely on a single device,
with no network transfers or offloading.

\begin{table}[ht]
\centering
\caption{Average per-inference latency and energy consumption for single-device execution.}
\label{tab:single_device_baseline}
\begin{tabular}{llll}
\hline
\textbf{Device\,\,} & \textbf{Model} & \textbf{Latency (ms)} & \textbf{Energy (J)} \\
\hline
\multirow{3}{*}{Edge}     & VGG16       & 666.870 & 8.002 \\
                               & AlexNet     & 132.400 & 1.589 \\
                               & MobileNetV2\,\, &  71.900 & 0.863 \\
\hline
\multirow{3}{*}{Fog} & VGG16       & 169.908 & 2.549 \\
                               & AlexNet     &  20.988 & 0.315 \\
                               & MobileNetV2\,\, &  15.954 & 0.239 \\
\hline
\multirow{3}{*}{Cloud}   & VGG16       &   1.164 & 0.037 \\
                               & AlexNet     &   0.830 & 0.024 \\
                               & MobileNetV2\,\, &   4.175 & 0.092 \\
\hline
\end{tabular}
\end{table}

Across all three models, the edge and cloud nodes differ by roughly two orders
of magnitude in both latency and energy. The cloud node completes VGG16 in
1.16\,ms and AlexNet in 0.83\,ms, while the edge node requires 666.87\,ms and
132.40\,ms respectively, a gap of approximately 570$\times$. MobileNetV2 shows
a considerably smaller gap of 17$\times$ (71.90\,ms vs 4.18\,ms), reflecting
its design for resource-constrained inference. The fog node sits between the
two extremes, with latencies of 169.91\,ms, 20.99\,ms, and 15.95\,ms for
VGG16, AlexNet, and MobileNetV2, respectively. The same ordering holds for
energy: the edge node consumes 0.86--8.00\,J per inference, the fog node
0.24--2.55\,J, and the cloud node 0.009--0.037\,J. These single-device figures
exclude any network transfer cost and therefore represent idealized lower bounds
rather than realistic end-to-end performance. In practice, routing all inference
to the cloud introduces network latency and bandwidth constraints that make
full offloading impractical in real IoT deployments, which is precisely the
motivation for adaptive partitioning.

\subsection{Static Partitioning Baseline}

Table~\ref{tab:static_partitioning_averages} reports the per-inference results
when the three models are executed across the three-node continuum using a fixed
partitioning scheme, with cut points chosen to distribute roughly one-third of
the feature extraction work to each node. For VGG16, this corresponds to layers
0--10 on the edge node, layers 11--30 on the fog node, and the classifier head
on the cloud node. For AlexNet, layers 0--9 run on the edge node, layers 10--12
together with the adaptive average pooling module on the fog node, and the
classifier on the cloud node. For MobileNetV2, the first 10 blocks run on the
edge node, the remaining 9 blocks on the fog node, and the pooling and
classifier head on the cloud node.

\begin{table}[H]
\caption{Per-inference latency and energy consumption under static partitioning.}
\centering
\renewcommand{\arraystretch}{1.2}
\begin{tabular}{llllll}
\toprule
\textbf{Model} & \textbf{Pipeline} & \textbf{Edge}  & \textbf{Fog} & \textbf{Cloud} & \textbf{Total} \\
               & \textbf{Latency (ms)}  & \textbf{Energy (J)} & \textbf{Energy (J)}   & \textbf{Energy (J)} & \textbf{Energy (J)} \\
\midrule
VGG16            & 525.142           & 2.297               & 2.491                 & 0.905               & 5.693 \\
AlexNet        & 78.148            & 0.237               & 0.082                 & 0.356               & 0.675 \\
MobileNetV2\,\,    & 98.457            & 0.624               & 0.268                 & 0.027               & 0.919 \\
\bottomrule
\end{tabular}
\label{tab:static_partitioning_averages}
\end{table}

Relative to the single-device baseline, static partitioning reduces end-to-end
latency for VGG16 by 21.2\% (from 666.87\,ms to 525.14\,ms) and for AlexNet
by 41.0\% (from 132.40\,ms to 78.15\,ms). MobileNetV2 behaves differently:
the network transfer overhead outweighs the computational benefit, increasing
end-to-end latency by 36.9\% from 71.90\,ms to 98.46\,ms. In terms of energy,
the total system cost is 5.69\,J for VGG16, 0.68\,J for AlexNet, and 0.92\,J
for MobileNetV2. The per-node energy distribution differs markedly across
models: for VGG16, the load is spread relatively evenly across all three nodes
(2.30\,J, 2.49\,J, and 0.91\,J for edge, fog, and cloud, respectively), while
for AlexNet and MobileNetV2 the edge node remains the dominant consumer
(0.24\,J and 0.62\,J), with the fog node contributing considerably less
(0.08\,J and 0.27\,J).

\subsection{Adaptive Partitioning}
Table~\ref{tab:adaptive_architecture_comparison} reports the same per-inference
metrics when the adaptive scheduler controls the split points. The scheduler
initializes by running 50 inferences at the static split to establish the
baseline threshold, followed by 15 inferences at each of three automatically
selected probe splits. It then fits the per-node compute rates and link
bandwidth parameters, selects the best starting split, and enters steady-state
operation with re-evaluation every 100 inferences.

\begin{table}[t]
\caption{Average per-inference latency and energy consumption under adaptive partitioning.}
\centering
\renewcommand{\arraystretch}{1.2}
\begin{tabular}{llllll}
\toprule
\textbf{Model} & \textbf{Pipeline} & \textbf{Edge}  & \textbf{Fog} & \textbf{Cloud} & \textbf{Total} \\
               & \textbf{Latency (ms)}  & \textbf{Energy (J)} & \textbf{Energy (J)}   & \textbf{Energy (J)} & \textbf{Energy (J)} \\
\midrule
VGG16         & 491.855           & 1.489               & 1.235                 & 0.930               & 3.654 \\
AlexNet        & 60.233            & 0.078               & 0.097                 & 0.259               & 0.434 \\
MobileNetV2\,\,    & 84.479            & 0.494               & 0.078                 & 0.099               & 0.670 \\
\bottomrule
\end{tabular}
\label{tab:adaptive_architecture_comparison}
\end{table}

Under adaptive partitioning, the total system energy is 3.65\,J, 0.43\,J, and
0.67\,J for VGG16, AlexNet, and MobileNetV2 respectively, with end-to-end
latencies of 491.86\,ms, 60.23\,ms, and 84.48\,ms. For all three models, the
adaptive framework reduces both energy consumption and latency relative to the
static baseline. The per-node energy distribution shifts noticeably between the
two configurations: for VGG16, the edge and fog node contributions drop from
2.30\,J to 1.49\,J and from 2.49\,J to 1.24\,J respectively. For AlexNet, the
edge node contribution drops from 0.24\,J to 0.08\,J. For MobileNetV2, the edge
node drops from 0.62\,J to 0.49\,J, while the cloud node rises from 0.03\,J
to 0.10\,J, reflecting a shift of heavier computation toward the more
energy-efficient cloud node.

\subsection{Comparison of Static and Adaptive Partitioning}
Table~\ref{tab:improvement_comparison} summarizes the improvements
achieved by the adaptive framework over the static baseline in terms of
percentage reductions in latency and total system energy.

\begin{table}[H]
\centering
\caption{Reduction in latency and energy achieved by adaptive over static partitioning.}
\label{tab:improvement_comparison}
\begin{tabular}{lll}
\hline
\textbf{Model} & \textbf{Latency Reduction (\%)} & \textbf{Energy Reduction (\%)} \\ \hline
VGG16      & 6.34  & 35.82 \\
AlexNet     & 22.92 & 35.70 \\
MobileNetV2\,\, & 14.20 & 27.09 \\ \hline
\end{tabular}
\end{table}

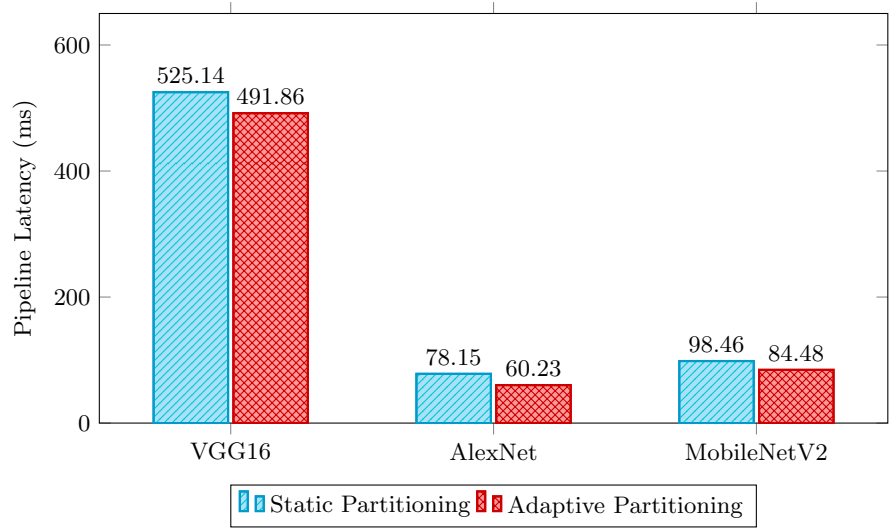
\begin{figure}[!t]
    \centering
    \begin{tikzpicture}
        \begin{axis}[
            ybar,
            bar width=28pt,
            width=12cm,
            height=6cm,
            enlarge x limits=0.25,
            ylabel={Pipeline Latency (ms)},
            symbolic x coords={VGG16, AlexNet, MobileNetV2},
            xtick=data,
            nodes near coords,
            nodes near coords align={vertical},
            every node near coord/.append style={font=\footnotesize},
            ymin=0, ymax=650, 
            legend style={at={(0.5,-0.15)}, anchor=north, legend columns=-1},
        ]
        
        \addplot[fill=cyan!30, postaction={pattern=north east lines, pattern color=cyan!80!black}, draw=cyan!80!black, thick] coordinates {
            (VGG16,525.14) 
            (AlexNet,78.15) 
            (MobileNetV2,98.46)
        };
        
        \addplot[fill=red!40, postaction={pattern=crosshatch, pattern color=red!80!black}, draw=red!80!black, thick] coordinates {
            (VGG16,491.86) 
            (AlexNet,60.23) 
            (MobileNetV2,84.48)
        };
        
        \legend{Static Partitioning, Adaptive Partitioning}
        \end{axis}
    \end{tikzpicture}
    \caption{Pipeline latency under static and adaptive partitioning.}
    \label{fig:latency_comparison}
\end{figure}

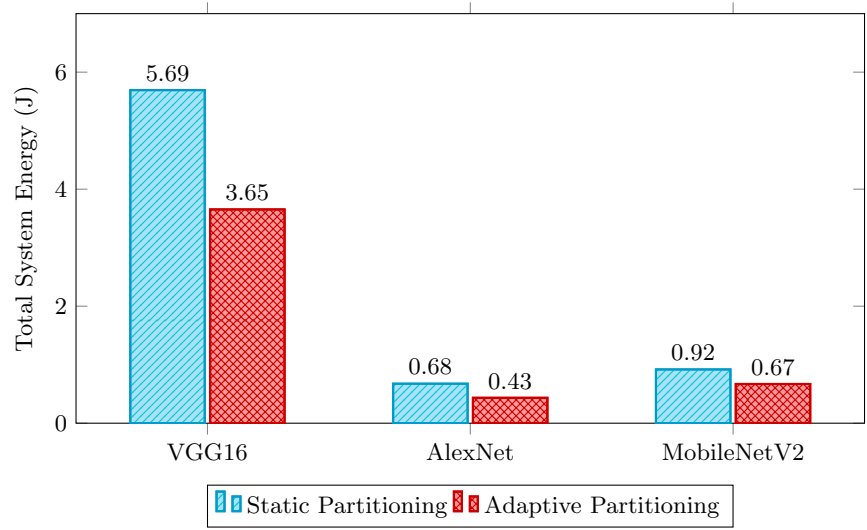
\begin{figure}[!t]
    \centering
    \begin{tikzpicture}
        \begin{axis}[
            ybar,
            bar width=28pt,
            width=12cm,
            height=6cm,
            enlarge x limits=0.25,
            ylabel={Total System Energy (J)},
            symbolic x coords={VGG16, AlexNet, MobileNetV2},
            xtick=data,
            nodes near coords,
            nodes near coords align={vertical},
            every node near coord/.append style={font=\footnotesize},
            ymin=0, ymax=7,
            legend style={at={(0.5,-0.15)}, anchor=north, legend columns=-1},
        ]
        
        \addplot[fill=cyan!30, postaction={pattern=north east lines, pattern color=cyan!80!black}, draw=cyan!80!black, thick] coordinates {
            (VGG16,5.693) 
            (AlexNet,0.675) 
            (MobileNetV2,0.919)
        };
        
        \addplot[fill=red!40, postaction={pattern=crosshatch, pattern color=red!80!black}, draw=red!80!black, thick] coordinates {
            (VGG16,3.654) 
            (AlexNet,0.434) 
            (MobileNetV2,0.670)
        };
        
        \legend{Static Partitioning, Adaptive Partitioning}
        \end{axis}
    \end{tikzpicture}
    \caption{Total system energy under static and adaptive partitioning.}
    \label{fig:energy_comparison}
\end{figure}

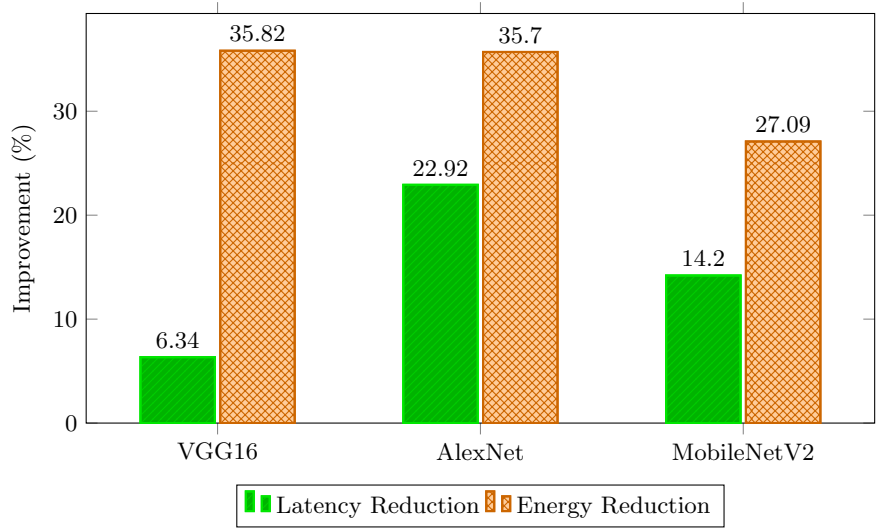
\begin{figure}[t]
\centering
\begin{tikzpicture}
\begin{axis}[
    ybar,
    bar width=28pt,
    width=12cm,
    height=7cm,
    ylabel={Improvement (\%)},
    symbolic x coords={VGG16,AlexNet,MobileNetV2},
    xtick=data,
    ymin=0,
    legend style={at={(0.5,-0.15)}, anchor=north, legend columns=-1},
    nodes near coords,
    enlarge x limits=0.25
]
\addplot[fill=green!70!black, draw=green!90!black, thick,
         postaction={pattern=north east lines, pattern color=green!80!black}]
    coordinates {(VGG16,6.34) (AlexNet,22.92) (MobileNetV2,14.20)};

\addplot[fill=orange!40, draw=orange!80!black, thick,
         postaction={pattern=crosshatch, pattern color=orange!80!black}]
    coordinates {(VGG16,35.82) (AlexNet,35.70) (MobileNetV2,27.09)};

\legend{Latency Reduction, Energy Reduction}
\end{axis}
\end{tikzpicture}
\caption{Reduction in latency and energy achieved by adaptive over static partitioning.}
\label{fig:improvement_percent}
\end{figure}

The energy reduction ranges from 27.09\% for MobileNetV2 to 35.82\% for VGG16,
and the latency reduction ranges from 6.34\% for VGG16 to 22.92\% for AlexNet.
Across all three models, the energy reduction consistently exceeds the latency
reduction, reflecting the framework's objective weighting. Figures
\ref{fig:latency_comparison} and \ref{fig:energy_comparison} show the absolute
latency and energy values side by side for both configurations, and
Fig.~\ref{fig:improvement_percent} shows the relative reductions as percentages.

The three models differ in how the savings are distributed between energy and
latency. VGG16 achieves the largest absolute and relative energy reduction
(2.04\,J, 35.82\%) but the smallest latency reduction (33.28\,ms, 6.34\%),
suggesting the scheduler prioritized offloading computation away from the
energy-hungry edge node over minimizing transfer overhead. AlexNet achieves both
a large energy reduction (0.24\,J, 35.70\%) and the largest latency reduction
of the three models (17.92\,ms, 22.92\%). MobileNetV2 shows the smallest energy
reduction (27.09\%) alongside a moderate latency improvement of 14.20\%. Across
all three models, energy reductions consistently exceed latency reductions in
percentage terms, consistent with the higher weights assigned to energy terms in
the objective function.

\vspace{1cm}

\section{Conclusions} \label{sec:conclusions}
This paper presented an adaptive partitioning and offloading framework for
distributed DNN inference across a heterogeneous edge-cloud continuum,
motivated by the lack of real-hardware evaluation and limited focus on energy
consumption in existing works. The framework was validated on a physical testbed
comprising a Raspberry Pi edge node, a laptop fog node, and a GPU-equipped cloud
node, across three CNN models spanning a broad complexity range. Results show
energy reductions of up to 35.82\% for VGG16 and AlexNet, and 27.09\% for
MobileNetV2, alongside latency reductions of up to 22.92\%, all relative to a
static partitioning baseline. These findings confirm that adaptive partitioning
can significantly reduce energy consumption on resource-constrained devices
without violating latency constraints, addressing a gap that simulation-based
approaches have largely been overlooked.

These results compare favorably with recent related work despite differences
in optimisation objective: Li et al.~\cite{li2025adaptive} achieved 27.31\%
latency reduction through joint partitioning and pruning, Chen et
al.~\cite{chen2024edgeci} reported 34.72\%--43.52\% latency improvements with
EdgeCI, and Fang et al.~\cite{fang2022ai} demonstrated substantial power
efficiency gains via DRL-based offloading. Our framework achieves comparable
improvements while targeting energy consumption over latency, and unlike
RL-based or parallel inference approaches, relies on a single weighted objective
function evaluated periodically at runtime, making it lightweight and
continuously adaptive.

\begin{credits}
\subsubsection{\ackname} : This work was partially supported by the MSCA's TALENTS (101299722) project. The computations were enabled by resources provided by the National Academic Infrastructure for Supercomputing in Sweden (NAISS), partially funded by the Swedish Research Council through grant agreement no. 2022-06725.
\subsubsection{\discintname}
The authors have no competing interests to declare that are
relevant to the content of this article.
\end{credits}
~\vspace{-10cm}
\bibliographystyle{splncs04}
\bibliography{ref}

\end{document}